\documentclass[11pt]{article}

\usepackage{amsmath,amsfonts,amsthm,amssymb}
\usepackage{graphicx,cite}
\usepackage[tight]{subfigure}
\usepackage[a4paper,margin=3cm]{geometry}
\usepackage{algorithm}
\usepackage{algpseudocode}

\newcommand{\ie}{i.\,e.}

\newcommand{\algorithmicinput}{\textbf{Input:}}
\newcommand{\algorithmicoutput}{\textbf{Output:}}
\newcommand{\Input}{\item[\algorithmicinput]}
\newcommand{\Output}{\item[\algorithmicoutput]}

\newtheorem{definition}{Definition}[section]
\newtheorem{corollary}[definition]{Corollary}
\newtheorem{theorem}[definition]{Theorem}
\newtheorem{lemma}[definition]{Lemma}

\newcommand{\bemph}[1]{\textbf{\boldmath #1}}

\newcommand{\match}{\ensuremath{\operatorname{Match}}}
\newcommand{\mst}{\ensuremath{\operatorname{MST}}}
\newcommand{\tsp}{\ensuremath{\operatorname{TSP}}}
\newcommand{\scc}{\ensuremath{\operatorname{SCC}}}
\newcommand{\acc}{\ensuremath{\operatorname{ACC}}}
\newcommand{\gfp}{\ensuremath{\operatorname{GFP}}}
\newcommand{\odd}{{\ensuremath{\operatorname{odd}}}}
\newcommand{\sol}{{\ensuremath{\operatorname{sol}}}}
\newcommand{\cc}{\ensuremath{\operatorname{CC}}}
\newcommand{\apx}{{\ensuremath{\operatorname{apx}}}}
\newcommand{\smst}{\ensuremath{\operatorname{MST}}}

\newcommand{\pareto}{\ensuremath{\mathcal{P}}}
\newcommand{\parapx}{\pareto^{\apx}}
\newcommand{\parcc}{\ensuremath{\mathcal{P}_{\cc}}}
\newcommand{\partsp}{\ensuremath{\mathcal{P}_{\tsp}}}
\newcommand{\parmst}{\ensuremath{\mathcal{P}_{\mst}}}
\newcommand{\parmatch}{\ensuremath{\mathcal{P}_{\match}}}

\newcommand{\class}[1]{\ensuremath{\mathsf{#1}}}
\newcommand{\NP}{\class{NP}}
\newcommand{\nat}{\ensuremath{\mathbb{N}}}

\newcommand{\mstsp}{\ensuremath{\Delta\operatorname{-STSP}}}

\newcommand{\gastsp}{\ensuremath{\Delta(\gamma)\operatorname{-STSP}}}
\newcommand{\matsp}{\ensuremath{\Delta\operatorname{-ATSP}}}
\newcommand{\gaatsp}{\ensuremath{\Delta(\gamma)\operatorname{-ATSP}}}
\newcommand{\stspot}{\ensuremath{\operatorname{STSP}(1,2)}}
\newcommand{\atspot}{\ensuremath{\operatorname{ATSP}(1,2)}}

\title{Approximation Algorithms for
Multi-Criteria~Traveling~Salesman~Problems%
\thanks{A preliminary version of this work has been presented at the 4th
Workshop on Approximation and Online Algorithms
(WAOA 2006)~\cite{MantheyRam:MultiCritTSP:2007}.}}

\author{Bodo Manthey%
\thanks{Supported by the Postdoc-Program of the German Academic Exchange Service
(DAAD). On leave from Saarland University. Work done in part at the Institute
for Theoretical Computer Science of the University of L\"ubeck supported by DFG
research grant RE 672/3 and at the Department of Computer Science at Saarland
University.}$^{~,1}$
\and
L.~Shankar Ram$^{2}$}

\date{\footnotesize \centering
$^1$ \parbox[t]{7.3cm}{Yale University \\ Department of Computer Science \\
P.~O.~Box 208285, New Haven, CT 06520-8285, USA \\
\texttt{manthey@cs.yale.edu}}  \qquad
$^2$ \parbox[t]{6.1cm}{ETH Z\"urich \\ Institut f\"ur Theoretische Informatik \\
8092 Z\"urich, Switzerland \\
\texttt{shankar.lakshminarayanan@ag.ch}}}

\begin{document}
\maketitle

\begin{abstract}
  We analyze approximation algorithms for several variants of the traveling
  salesman problem with multiple objective functions. First, we consider the
  symmetric TSP (STSP) with $\gamma$-triangle inequality. For this problem, we
  present a deterministic polynomial-time algorithm that achieves an
  approximation ratio of $\min\{1+\gamma,
  \frac{2 \gamma^2}{2 \gamma^2 - 2 \gamma +1}\} + \varepsilon$ and a randomized
  approximation algorithm that achieves a ratio of
  $\frac{2\gamma^3 + 2 \gamma^2}{3\gamma^2 - 2\gamma +1}  + \varepsilon$. In
  particular, we obtain a $2+\varepsilon$ approximation for multi-criteria
  metric STSP.

  Then we show that multi-criteria cycle cover problems admit fully
  polynomial-time randomized approximation schemes. Based on these schemes, we
  present randomized approximation algorithms for STSP with $\gamma$-triangle
  inequality (ratio $\frac{1+\gamma}{1+3 \gamma - 4 \gamma^2} + \varepsilon$),
  asymmetric TSP (ATSP) with $\gamma$-triangle inequality (ratio $\frac 12 +
  \frac{\gamma^3}{1-3\gamma^2} + \varepsilon$), STSP with weights one and two
  (ratio $4/3$) and ATSP with weights one and two (ratio $3/2$).
\end{abstract}

%%%%%%%%%%%%%%%%%%%%%%
\section{Introduction}
%%%%%%%%%%%%%%%%%%%%%%

In many practical optimization problems, there is not only one single objective
function to measure the quality of a solution, but there are several such
functions. Consider for instance buying a car: We (probably) want to buy a cheap
car that is fast and has a good gas mileage. How do we decide which car is the
best one for us? Of course, with respect to any single criterion, making the
decision is easy. But with multiple criteria involved, there is no natural
notion of a best choice. The aim of \emph{multi-criteria optimization} (also
called multi-objective optimization or Pareto optimization) is to cope with this
problem. To transfer the concept of a best choice to multi-criteria
optimization, the notion of \emph{Pareto curves} was introduced (cf.\
Section~\ref{ssec:prelim} and Ehrgott~\cite{Ehrgott:MulticriteriaOpt:2005}). A
Pareto curve is a set of solutions that can be considered optimal.

However, for most optimization problems, Pareto curves cannot be computed
efficiently. Thus, we have to be content with approximations to them.

The \emph{traveling salesman problem} (TSP) is one of the best-known
combinatorial optimization
problems~\cite{GutinPunnen:TSP:2002,LawlerEA:TSP:1985}. An instance of the TSP
is a complete graph with edge weights, and the aim is to find a Hamiltonian
cycle (also called a tour) of minimum weight. Since the TSP is
\NP-hard~\cite{GareyJohnson:NP:1979}, we cannot hope to always find an optimal
tour efficiently. For practical purposes, however, it is often sufficient to
obtain a tour that is close to optimal. In such cases, we require
\emph{approximation algorithms}, \ie, polynomial-time algorithms that compute
such near-optimal tours.

While the approximability of several variants of the single-criterion TSP has
been studied extensively in the past decades, not much is known about the
approximability of multi-criteria TSP. The classical TSP is about a traveling
salesman who has to visit a certain number of cities and return back home in a
shortest tour. ``Real'' saleswomen and salesmen do not face such a simple
situation. Instead, while arranging their tours, they have to bear in mind
several objectives that are to be optimized. For instance, the distance
traveled and the travel time should be minimized while the journey should be as
cheap as possible. This gives rise to multi-criteria TSP, for which we design
approximation algorithms in this paper.

%%%%%%%%%%%%%%%%%%%%%%%%%%
\subsection{Preliminaries}
\label{ssec:prelim}

\paragraph{Graphs and Optimization Problems.}

Let $G = (V,E)$ be a graph (directed or undirected) with edge weights
$w: E \rightarrow \nat$. We define the weight of a subgraph $G'=(V', E')$ of $G$
or a subset $E'$ of the edges of $G$ as the sum of the weights of its edges:
$w(G') = w(E') = \sum_{e \in E'}w(e)$. For $k \in \nat$, we define
$[k] = \{1,2, \ldots, k\}$.

TSP in general is the following optimization problem: Given a graph with edge
weights, find a Hamiltonian cycle, \ie, a cycle that visits every vertex of the
graph exactly once, of minimum weight. In case of undirected graphs, we speak of
the symmetric TSP (STSP), while in case of directed graphs, we refer to the
problem as the asymmetric TSP (ATSP).

An instance of \bemph{\mstsp} is an undirected complete graph $G=(V,E)$ with
edge weights $w: E \rightarrow \nat$ that fulfill triangle inequality, \ie,
$w(\{u,v\}) \leq w(\{u,x\})+ w(\{x,v\})$ for all distinct vertices
$u, v, x\in V$.

For $\gamma \in [\frac 12, 1]$, \bemph{\gastsp} is the restriction of \mstsp\ to
instances that satisfy $\gamma$-strengthened triangle inequality, \ie,
$w(\{u,v\}) \leq \gamma \cdot (w(\{u,x\})+ w(\{x,v\}))$ for all distinct
vertices $u, v, x$.

\bemph{\stspot} is the special case of \mstsp\ where only one and two are
allowed as edge weights, \ie, $w: E \rightarrow \{1,2\}$.

\bemph{\matsp}, \bemph{\gaatsp}, and \bemph{\atspot} are defined like their
undirected counterparts \mstsp, \gastsp, and \stspot, respectively, except that
the graphs are directed.

Note that for $\gamma = 1$, \gastsp\ and \gaatsp\ become \mstsp\ and \matsp,
respectively. As $\gamma$ gets smaller, the edge weights become more and more
structured. For $\gamma = 1/2$, all edge weights are equal. The
$\gamma$-strengthened triangle inequality can also be considered as a
data-dependent bound~\cite{BlaeserEA:ATSP:2006}: Given an instance of metric
TSP, we compute the minimum $\gamma$ such that the instance fulfills
$\gamma$-strengthened triangle inequality. If $\gamma < 1$, then we obtain a
better performance guarantee for our approximate solution than with triangle
inequality alone.

A cycle cover of a graph $G=(V,E)$ is a subgraph $(V,C)$ that consists solely of
cycles such that every vertex $v \in V$ is part of exactly one cycle. In most
cases, we refer to a cycle cover as the set $C$ of its edges. Hamiltonian cycles
are cycle covers that consist of only a single cycle.

The problem of computing cycle covers of minimum weight in undirected graphs is
called \bemph{\scc}. The directed version of the problem is called \bemph{\acc}.

\paragraph{Multi-Criteria Optimization.}

A $k$-criteria optimization problem consists of a set $I$ of instances, a set
$\sol(x)$ of feasible solutions for every instance $x \in I$, $k$ objective
functions $w_1, \ldots, w_k$, each mapping pairs of $x \in I$ and
$y \in \sol(x)$ to $\nat$, and $k$ types indicating whether $w_i$ should be
minimized or maximized. We refer to Ehrgott and
Gandibleux~\cite{Ehrgott:MulticriteriaOpt:2005,%
   EhrgottGandibleux:Multiobjective:2000} for surveys on multi-criteria
optimization problems. Throughout this paper, we restrict ourselves to problems
where all objective functions should be minimized. Furthermore, we assume that
the number $k$ of criteria is fixed. The running-times of our algorithms are
exponential in $k$. But since $k$ is typically a small number, this does not
cause any harm.

The optimization problems defined in Section~\ref{ssec:prelim} are generalized
to their multi-criteria counterparts in the obvious way: We have $k$ objective
functions $w_1, \ldots, w_k$, each induced by edge weight functions (to which we
also refer as $w_1, \ldots, w_k$) as described. If we have additional
restrictions on the edge weights, like the triangle inequality, every edge
weight function is assumed to fulfill them.

In general, the different objective functions are in conflict with each other,
\ie, it is impossible to minimize all of them simultaneously. Therefore, the
notion of Pareto curves has been introduced. For the following definitions, let
$\Pi$ be a $k$-criteria optimization problem as defined above.

A set $\pareto(x) \subseteq \sol(x)$ is called a \bemph{Pareto curve} of $x$ if
for all solutions $z \in \sol(x)$, there exists a solution $y \in \pareto(x)$
with $w_i(x,y) \leq w_i(x,z)$ for all $i \in [k]$.

A Pareto curve contains all solutions that might be considered optimal. If there
are two solutions $y$ and $z$ with $w_i(x,y) = w_i(x,z)$ for all $i \in [k]$,
then it suffices to put one of them into $\pareto(x)$. For completeness, let us
mention that Pareto curves are not unique in general: In our definition, it is
not forbidden to include dominated solutions in $\pareto(x)$ (a solution $y$ is
dominated if there exists a $z$ with $w_i(x,z) \leq w_i(x,y)$ for all
$i \in [k]$ and $w_i(x,z) < w_i(x,y)$ for some $i \in [k]$, \ie, $z$ is strictly
better than $y$).

For the majority of multi-criteria problems, computing Pareto curves is hard for
two reasons: First, many two-criteria problems allow for a reduction from the
knapsack problem. Second, Pareto curves are often of exponential size.
Therefore, we have to be content with approximate Pareto curves. Let
$\beta \geq 1$, and let $x \in I$ and $\parapx(x) \subseteq \sol(x)$. The set
$\parapx(x)$ is called a \bemph{$\beta$-approximate Pareto curve} for $x$ if,
for every $z \in \sol(x)$, there exists a $y \in \parapx(x)$ with
$w_i(x,y) \leq \beta \cdot w_i(x,z)$ for all $i \in [k]$.

A $1$-approximate Pareto curve is a Pareto curve. For completeness, let us
mention that if $\Pi$ is a maximization problem (or an objective $w_i$ for some
$i \in [k]$ should be maximized), then the condition is
$w_i(x,z) \leq \beta \cdot w_i(x,y)$.

While Pareto curves itself are often of exponential size, it is known that
$(1+\varepsilon)$-approximate Pareto curves of size polynomial in the input size
and $1/\varepsilon$ exist~\cite{PapadimitriouYannakakis:TradeOffs:2000}. (The
technical restriction is that the objective functions are restricted to assume
values of at most $2^{p(|x|)}$ for $x\in I$ and some polynomial $p$.)

The above definition leads immediately to the notion of an approximation
algorithm for multi-criteria optimization problems: Let $\beta \geq 1$. A
\bemph{$\beta$-approximation algorithm} for $\Pi$ is an algorithm that, for
every input $x \in I$, computes a $\beta$-approximate Pareto curve for $x$ in
time polynomial in the size $|x|$ of $x$.

A \bemph{randomized $\beta$-approximation algorithm} for $\Pi$ is a
polynomial-time algorithm that, for every input $x \in I$, computes a set
$\parapx(x) \subseteq \sol(x)$ such that $\parapx(x)$ is a $\beta$-approximate
Pareto curve for $x$ with a probability of at least $1/2$.

By executing a randomized approximation algorithm $\ell$ times, we obtain a
$\beta$-appro\-ximate Pareto curve with a probability of at least
$1- 2^{-\ell}$, \ie, the failure probability tends exponentially to zero: We
take the union of all sets of solutions computed in the $\ell$ iterations and
throw away all solutions that are dominated by solutions in the union.

Given the notion of (randomized) approximation algorithms, we can define
approximation schemes. A \bemph{fully polynomial-time approximation scheme
(FPTAS)} for $\Pi$ is an algorithm that, on input $x \in I$ and
$\varepsilon > 0$, computes a $(1+\varepsilon)$-approximate Pareto curve in time
polynomial in the size of $x$ and~$1/\varepsilon$.

A \bemph{fully polynomial-time randomized approximation scheme (\mbox{FPRAS})}
for $\Pi$ is a randomized approximation algorithm that, on input $x \in I$ and
$\varepsilon > 0$, computes a $(1+\varepsilon)$-approximate Pareto curve in time
polynomial in the size of $x$ and $1/\varepsilon$.

Finally, we define the notion of a randomized exact algorithm: A
\bemph{randomized exact algorithm} for $\Pi$ is an algorithm that, on input $x$,
computes a Pareto curve of $x$ in time polynomial in the size of $x$ with a
probability of at least $1/2$.

An optimization problem $\Pi$ is said to be polynomially bounded if there exists
a polynomial $p$ such that the following holds for every objective function
$w_i$ of $\Pi$: For every instance $x$ and every feasible solution $y$ for $x$,
$w_i(x,y) \leq p(|x|)$ for all $i \in [k]$. Analogously to the fact that a
polynomially bounded single-criterion problem that admits an FPTAS can be solved
exactly in polynomial time (cf.\ Ausiello et
al.~\cite[Theorem~3.15]{AusielloEA:ComplApprox:1999}), randomized exact
algorithms exist for polynomially bounded multi-criteria optimization problems
that admit an FPRAS.

%%%%%%%%%%%%%%%%%%%%%%%%%%%%%
\subsection{Previous Results}
\label{ssec:previous}

The approximability of single-criterion TSP has been studied intensively in the
past. Table~\ref{tab:tsp} shows the currently best approximation ratios of the
variants  for which the multi-criteria counterparts are considered in this
paper.

\begin{table}[t]
\centering 
\begin{tabular}{|l|c|l|} \hline 
\textbf{Variant} & \textbf{Ratio} & \textbf{Reference} \\ \hline 
\mstsp & $3/2$ & Christofides~\cite{Christofides:TSP:1976} \\ \hline
\gastsp &
$\min\bigl\{\frac{3\gamma^2}{3\gamma^2-2\gamma+1},
\frac{2-\gamma}{3-3\gamma}\bigr\}$ &
B\"ockenhauer et al.~\cite{BoeckenhauerEA:SharpenedIPL:2000} \\ \hline 
\stspot & $8/7$ &
Berman, Karpinski~\cite{BermanKarpinski:87TSP12:2006} \\ \hline 
\matsp & $0.842 \cdot \log n$ &
Kaplan et al.~\cite{KaplanEA:TSP:2005} \\ \hline
\gaatsp & $\min\bigl\{\frac{1+\gamma}{2-\gamma- \gamma^3}, 
\frac{\gamma}{1-\gamma}\bigr\}$ & Bl\"aser et al.~\cite{BlaeserEA:ATSP:2006};
Chandran and Ram~\cite{ChandranRam:Parameterized:2007} \\ \hline
\atspot & $5/4$ & Bl\"aser~\cite{Blaeser:ATSPZeroOne:2004} \\ \hline 
\end{tabular}
\caption{Approximability of single-criterion TSP.}
\label{tab:tsp}
\end{table}

While single-criterion optimization problems and their approximation properties
have been the subject of a considerable amount of research (cf.\ Ausiello et
al.~\cite{AusielloEA:ComplApprox:1999} for a survey), not much is known about
the approximability of multi-criteria optimization problems.

Papadimitriou and Yannakakis~\cite{PapadimitriouYannakakis:TradeOffs:2000}, by
applying results of Barahona and
Pulleyblank~\cite{BarahonaPulleyBlank:Arborescences:1987}, Mulmuley et
al.~\cite{MulmuleyEA:Matching:1987}, and
themselves~\cite{PapadimitriouYannakakis:RestrictedTree:1982}, showed that there
exist FPTASs for multi-criteria minimum-weight spanning trees and the
multi-criteria shortest path problem and an FPRAS (more precisely, a fully
polynomial \class{RNC} scheme) for the multi-criteria minimum weight matching
problem. The results were established by showing that a multi-criteria problem
admits an FPTAS if the exact version of the single-criterion problem can be
solved in pseudo-polynomial time. Let $\Pi$ be a single-criterion optimization
problem with instance set $I$ and objective function $w$. The
\bemph{exact version of $\Pi$} is the following decision problem: Given an
instance $x \in I$ and a number $W \in \nat$, does there exist a solution
$y \in \sol(x)$ with $w(x,y) = W$?

The exact versions of many single-criterion optimization problems are
\NP-complete since knapsack can be reduced to them easily. But this does not
rule out the possibility of pseudo-polynomial-time algorithms for them.

Multi-criteria TSP has been investigated by Ehrgott~\cite{Ehrgott:MultiTSP:2000}
and Angel et al.~\cite{AngelEA:BicritTSP:2004,AngelEA:MultiTSP:2005}.
Ehrgott~\cite{Ehrgott:MultiTSP:2000} considered a generalization of
Christofides' algorithm for \mstsp. Instead of considering Pareto curves, he
measured the quality of a solution $y$ for an instance $x$ as a norm of the
vector $(w_1(x,y), \ldots, w_k(x,y))$. Thus, he encoded the different objective
functions into a single one, which reduces the problem to a single-criterion
problem. The approximation ratio achieved is between $3/2$ and $2$, depending on
the norm used to combine the different criteria. However, by encoding all
objective functions into a single one, we lose the special properties of
multi-criteria optimization problems.

Angel et al.~\cite{AngelEA:BicritTSP:2004} considered two-criteria \stspot. They
presented a $3/2$-approxi\-mation algorithm for this problem by using a local
search heuristic. Finally, Angel et al.~\cite{AngelEA:MultiTSP:2005} generalized
these results to $k$-criteria \stspot\ by presenting a
$2-\frac{2}{k+1}$-approximation for $k\ge 3$. Although for every fixed $k$, the
approximation ratio is below $2$, it converges to $2$ as $k$ increases. Thus,
the ratio tends to the trivial ratio of 2, which can be achieved by selecting
any Hamiltonian cycle. These two are the only papers about the approximability
of Pareto curves of multi-criteria TSP we are aware of.

%%%%%%%%%%%%%%%%%%%%%%%%
\subsection{Our Results}

All our results hold for  an arbitrary but fixed number of objective functions.

We present a deterministic polynomial-time algorithm that computes
$(2 + \varepsilon)$-approxi\-mate Pareto curves for \mstsp\
(Section~\ref{subsec:det2}). This is the first efficient algorithm for computing
approximate Pareto curves for this problem. In fact, we show the following more
general result: If the edge weights satisfy $\gamma$-strengthened triangle
inequality for $\gamma \in [\frac 12, 1]$, then the algorithm computes a
$(\min\{1+\gamma, \frac{2 \gamma^2}{2 \gamma^2 - 2 \gamma +1}\} +
\varepsilon)$-approximate Pareto curve for arbitrarily small $\varepsilon > 0$
in polynomial time.

We generalize Christofides' algorithm~\cite{Christofides:TSP:1976} (cf.\
Vazirani~\cite[Sect.~3.2]{Vazirani:ApproxAlg:2001}) to obtain a randomized
approximation algorithm for multi-criteria \gastsp\
(Section~\ref{subsec:christofides}). For $\gamma \in [\frac 12, 1]$, our
algorithm achieves an approximation performance of
$\frac{2\gamma^3 + 2 \gamma^2}{3\gamma^2 - 2\gamma +1} + \varepsilon$. For
$\gamma=1$, this yields a ratio of $2 + \varepsilon$.

We consider cycle covers in Section~\ref{sec:cycles}. Cycle covers play an
important role in the design of approximation algorithms for the TSP. We prove
that there exists an FPRAS for computing approximate Pareto curves of
multi-criteria cycle covers. Subsequently, we extend this result and show that
the multi-criteria variant of the problem of finding graph factors of minimum
weight admits an FPRAS, too.

Finally, we analyze a randomized cycle-cover-based algorithm for multi-criteria
TSP (Section~\ref{sec:ccalgo}): We start by computing an approximate Pareto
curve of cycle covers. Then, for every cycle cover in the set computed, we
remove one edge of every cycle and join the paths thus obtained to a Hamiltonian
cycle. We analyze the approximation ratio of this algorithm for \gastsp\
(Section~\ref{subsec:ccgastsp}, approximation ratio
$\frac{1+\gamma}{1+3 \gamma - 4 \gamma^2} + \varepsilon$ for $\gamma < 1$),
\gaatsp\ (Section~\ref{subsec:ccgaatsp}, ratio
$\frac 12  +  \frac{\gamma^3}{1 - 3 \gamma^2}  + \varepsilon$ for
$\gamma < 1/\sqrt 3$), \stspot, and \atspot\ (Section~\ref{subsec:tspot}, ratios
$4/3$ and $3/2$, respectively).

As far as we know, our algorithms are the first approximation algorithms for
Pareto curves for \mstsp, \gastsp, \gaatsp, and \atspot. Furthermore, we achieve
a better approximation ratio for \stspot\ than the approximation algorithms by
Angel et al.~\cite{AngelEA:BicritTSP:2004,AngelEA:MultiTSP:2005} for all $k$.

%%%%%%%%%%%%%%%%%%%%%%%%
\section{Metric TSP}
\label{sec:christofides}
%%%%%%%%%%%%%%%%%%%%%%%%

In this section, we present two algorithms for \mstsp\ and \gastsp. Another
approximation algorithm that can be used for approximating \gastsp, which is
based on computing cycle covers, will be presented in Section~\ref{sec:ccalgo}.

The analyses of the algorithms in this section exploit the following result due
to B\"ockenhauer et al.~\cite{BoeckenhauerEA:SharpenedIPL:2000}.

\begin{lemma}[B\"ockenhauer et al.~\cite{BoeckenhauerEA:SharpenedIPL:2000}]
\label{lem:gastsp}
  Let $G=(V,E)$ be an undirected complete graph with an edge weight function $w$
  satisfying $\gamma$-strengthened triangle inequality for some
  $\gamma \in [\frac 12, 1)$.

  Let $w_{\max} = \max_{e \in E}(w(e))$ and $w_{\min} = \min_{e \in E}(w(e))$ be
  the weights of a heaviest and lightest edge, respectively. Then
  $\frac{w_{\max}}{w_{\min}} \leq \frac{2\gamma^2}{1-\gamma}$.

  Let $e$ and $e'$ be two edges with a common endpoint. Then
  $\frac{w(e)}{w(e')} \leq \frac{\gamma}{1-\gamma}$.
\end{lemma}

Furthermore, we observe the following: Omitting two edges by taking a shortcut
reduces the weight by at least $2 \cdot (1-\gamma) \cdot w_{\min}$: The reason
is that the two edges $(u, v)$ and $(v,x)$ are replaced by $(u,x)$ and
$w(u,x) \leq \gamma \cdot (w(u,v) + w(v,x))$. Thus, the weight is reduced by at
least $w(u,v) + w(v,x) - w(u,x) \geq (1-\gamma) \cdot (w(u,v) + w(v,x)) \geq
2 \cdot (1-\gamma) \cdot w_{\min}$.

%%%%%%%%%%%%%%%%%%%%%%%%%%%%%%%%%%%%%%%%%%%%%%%%%%%%
\subsection{The Generalized Tree Doubling Algorithm}
\label{subsec:det2}

Consider the following approximation algorithm for single-criterion \mstsp,
which was first analyzed by Rosenkrantz et
al.~\cite{RosenkrantzEA:AnalysisHeuristicsTSP:1977} (cf.\
Vazirani\cite[Sect.~3.2]{Vazirani:ApproxAlg:2001}): First, we compute a minimum
spanning tree. Then we duplicate each edge. The result is an Eulerian graph. We
obtain a Hamiltonian cycle from this graph by walking along an Eulerian cycle.
If we come back to a vertex that we have already visited, we omit it and take a
short-cut to the next vertex in the Eulerian cycle. In this way, we obtain an
approximation ratio of $2$ for single-criterion \mstsp.
Algorithm~\ref{Algo:Tree} is an adaptation of this algorithm to multi-criteria
STSP. In the following, we estimate the approximation performance of this
algorithm.

\begin{algorithm}[t]
\begin{algorithmic}[1]
\Input undirected complete graph $G=(V,E)$; $k$ edge weight functions $w_i:
E \rightarrow \nat$ ($i \in [k]$); $\varepsilon > 0$
\Output an approximate Pareto curve $\partsp^{\apx}$ to the multi-criteria STSP
\State compute a $(1+\frac{\varepsilon}2)$-approximate Pareto curve
$\parmst^{\apx}$ for \smst\ on $G$ using the algorithm by Papadimitriou and
Yannakakis~\cite{PapadimitriouYannakakis:TradeOffs:2000}
\ForAll{trees $T \in \parmst^{\apx}$}
\State duplicate all edges in $T$ to obtain an Eulerian graph $\tilde T$
\State obtain a Hamiltonian cycle $S$ from $\tilde T$ by taking shortcuts
\State put $S$ into $\partsp^{\apx}$
\EndFor
\end{algorithmic}
\caption{The tree doubling algorithm for multi-criteria \mstsp.}
\label{Algo:Tree}
\end{algorithm}

\begin{theorem}
\label{thm:mstgamma}
  For all $\gamma \in [\frac 12, 1]$, Algorithm~\ref{Algo:Tree} computes a
  $(\min\{1 + \gamma, \frac{2 \gamma^2}{2 \gamma^2 - 2 \gamma + 1} \}
  + \varepsilon)$-approximate Pareto curve for multi-criteria \gastsp\ in time
  polynomial in the input size and~$1/\varepsilon$.
\end{theorem}

\begin{proof}
  We present two analyses showing approximation ratios of
  $1+\gamma + \varepsilon$ and $\frac{2 \gamma^2}{2 \gamma^2 - 2 \gamma + 1} +
  \varepsilon$, respectively. The first analysis holds for
  $\gamma \in [\frac 12, 1]$ while the second one only holds for
  $\gamma \in [\frac 12,1)$. However, $1+\gamma =
  \frac{2 \gamma^2}{2 \gamma^2 - 2 \gamma + 1}$ for $\gamma = 1$.

  The key observation of the first analysis is the following: Let
  $T \in \parmst^\apx$, and let $e$ be any edge in $T$. Then $e$ appears twice
  in $\tilde T$, but $e$ cannot appear twice in $S$ since $S$ is a Hamiltonian
  cycle. (We assume that $G$ contains at least three vertices.) Thus, at least
  one copy of $e$ is omitted. This is the moment at which the strengthened
  triangle inequality comes into play. Let $e_1, e_2, \ldots, e_\ell$ with
  $e_j = \{v_{j-1}, v_j\}$ be a path along the Eulerian cycle in $\tilde T$ such
  that this path is replaced by the edge $\{v_0, v_\ell\}$ by taking a shortcut.
  Then we have 
  \[
         w_i(\{v_0, v_\ell\})
    \leq \gamma \cdot \bigl(w_i(e_1) + w_i(e_2) + \ldots +  w_i(e_\ell)\bigr)
  \]
  by iteratively applying $\gamma$-strengthened triangle inequality. (We exploit
  the fact that $\gamma^c \leq \gamma$ for all $c \geq 1$.) Overall, every edge
  that we omit contributes at most a fraction of $\gamma$ of its weight. Since
  we omit at least one copy of every edge $e$, the two copies of $e$ contribute
  at most $(1+\gamma) \cdot w_i(e)$ to $S$. Thus,
  \[
    w_i(S) \leq (1 + \gamma) \cdot w_i(T)
  \]
  for all $i \in [k]$.

  To estimate the overall approximation performance, let $S'$ be an arbitrary
  Hamiltonian cycle. By omitting one edge, we obtain a tree $T'$. Since
  $\parmst^\apx$ is a $(1+\varepsilon/2)$-approximate Pareto curve for
  multi-criteria minimum-weight spanning trees on $G$, there exists a tree
  $T \in \parmst^\apx$ with
  \[
         w_i(T)
    \leq \bigl(1+\frac{\varepsilon}2\bigr) \cdot w_i(T')
    \leq \bigl(1+\frac{\varepsilon}2\bigr) \cdot w_i(S')
  \]
  for all $i \in [k]$. Let $S$ be the Hamiltonian cycle obtained from $T$, then
  \[
         w_i(S)
    \leq (1+\gamma) \cdot  w_i(T)
    \leq (1 + \gamma + \varepsilon) \cdot w_i(S')
  \]
  for all $i \in [k]$.

  For the second analysis, let again $S'$ be an arbitrary Hamiltonian cycle.
  This analysis only holds for $\gamma < 1$ since $w_{\max}/w_{\min}$ can be
  unbounded for $\gamma = 1$. All arguments hold simultaneously for all criteria
  $i \in [k]$. Without loss of generality, we assume that
  $\min_{e \in E} w_i(e) =1$ for all $i \in [k]$, \ie, $w_{\min} = 1$. By
  removing one edge of $S'$, we obtain a tree with a weight of at most
  $w_i(S') - w_{\min} = w_i(S') - 1$. Thus, there exists a tree
  $T \in \parmst^\apx$ from which we obtain a Eulerian graph $\tilde T$ with
  \[
         w_i(\tilde T)
    \leq 2 \cdot \bigl(1+ \frac{\varepsilon}2\bigr) \cdot (w_i(S') -1) 
    =    (2+\varepsilon) \cdot (w_i(S') -1).
  \]
  Let $n = |V|$ be the number of vertices of the whole graph. Then $\tilde T$
  contains $2n -2$ edges. Thus, in order to obtain a Hamiltonian cycle, we have
  to remove $n-2$ edges by taking shortcuts. Every shortcut decreases the weight
  by at least $2 \cdot (1-\gamma)$. Hence,
  \begin{eqnarray*}
  w_i(S) &\leq & w_i(\tilde T) - (n-2) \cdot 2 \cdot (1-\gamma) \\
         &\leq & (2+\varepsilon) \cdot w_i(S') - (2+\varepsilon)
               - (n-2) \cdot 2 \cdot (1-\gamma) \\
         &\leq & (2+\varepsilon) \cdot w_i(S') - 2 n \cdot (1-\gamma)
               - (4 \gamma -2) \\
         &\leq & (2+\varepsilon) \cdot w_i(S') - 2 n \cdot (1-\gamma)
  \end{eqnarray*}
  since $2 \gamma \geq 1$. We have $w_i(S') =
  \frac{2 n \gamma^2}{(1-\gamma) \cdot \alpha}$ for some appropriately chosen
  $\alpha \geq 1$, which implies
  \[
         \frac{w_i(S)}{w_i(S')}
    \leq 2+\varepsilon - \frac{2n \cdot (1-\gamma)^2 \cdot \alpha}{2 \gamma^2 n}
    =    2+\varepsilon - \frac{(1-\gamma)^2 \cdot \alpha}{\gamma^2} .
  \]
  Since $w_i(S) \leq n \frac{2\gamma^2}{1-\gamma}$, we also have
  $\frac{w_i(S)}{w_i(S')}  \leq \alpha$. Thus, the approximation ratio achieved
  is
  \begin{eqnarray*}
    &     & \max_{\alpha\geq 1}\left(\min\left(\alpha, 2+\varepsilon -
       \frac{(1-\gamma)^2 \cdot \alpha}{\gamma^2} \right)\right) \\
    &\leq & \max_{\alpha\geq 1}\left(\min\left(\alpha, 2 - \frac{(1-\gamma)^2
       \cdot \alpha}{\gamma^2} \right)\right) + \varepsilon \\
    &  =  & \frac{2 \gamma^2}{2 \gamma^2 - 2 \gamma +1} + \varepsilon,
  \end{eqnarray*}
  which completes the proof of the theorem.
\end{proof}

For small values of $\gamma$, the bound of
$\frac{2 \gamma^2}{2 \gamma^2 - 2 \gamma +1} + \varepsilon$ is the stronger one,
while $1+\gamma + \varepsilon$ yields a better bound in case of
$\gamma > 1/\sqrt 2$.

\begin{corollary}
\label{cor:mst}
Algorithm~\ref{Algo:Tree} computes $(2+\varepsilon)$-approximate Pareto curves
for multi-criteria \mstsp\ in time polynomial in the input size and
$1/\varepsilon$.
\end{corollary}

%%%%%%%%%%%%%%%%%%%%%%%%%%%%%%%%%%%%%%%%%%%%%%%%%%%%%%%%
\subsection{A Generalization of Christofides' Algorithm}
\label{subsec:christofides}

In this section, we generalize Christofides' algorithm to multi-criteria \mstsp,
which is the best approximation algorithm for single-criterion \mstsp\ known so
far. This algorithm computes approximate Pareto curves of matchings. In case of
single-criterion \mstsp, we can always find a matching with a weight of at most
half of the weight of the optimal Hamiltonian cycle. This is in contrast to
multi-criteria \mstsp, where the weights of the matchings can be arbitrarily
close to the weight of the optimal Hamiltonian cycle. The reason is that we
cannot choose the lighter of two different matchings since multiple objective
functions are involved; the term "`lighter"' is not well defined. Therefore, we
only get an approximation ratio of roughly two in this case. But for \gastsp, we
can show a better upper bound.

\begin{algorithm}[t]
\begin{algorithmic}[1]
\Input undirected complete graph $G=(V,E)$; $k$ edge weight functions
   $w_i: E \rightarrow \nat$ ($i \in [k]$); $\varepsilon > 0$
\Output an approximate Pareto curve $\partsp^{\apx}$ to the multi-criteria STSP
   (with a probability of at least $1/2$)
\State compute a $(1+\frac{\varepsilon}2)$-approximate Pareto curve
   $\parmst^{\apx}$ for \smst\ on $G$ using the algorithm by Papadimitriou and
   Yannakakis~\cite{PapadimitriouYannakakis:TradeOffs:2000}
\State let $p$ be the number of trees in $\parmst^{\apx}$ 
\ForAll{trees $T \in \parmst^{\apx}$}
\State let $V_\odd \subseteq V$ be the set of vertices of odd degree in $T$
\State compute $\parmatch^{\apx}(T)$ such that $\parmatch^{\apx}(T)$ is a
   $\bigl(1+\frac{\varepsilon}2\bigr)$-approximate Pareto curve for the
   minimum-weight matching problem on the complete graph induced by $V_\odd$
   with a probability of at least $1-\frac{1}{2p}$ using the algorithm by
   Papadimitriou and Yannakakis~\cite{PapadimitriouYannakakis:TradeOffs:2000}
\ForAll{matchings $M \in \parmatch^{\apx}(T)$}
\State let $S$ be a Hamiltonian cycle obtained from $T \cup M$ by taking shortcuts
\State put $S$ into $\partsp^{\apx}$
\EndFor
\EndFor
\end{algorithmic}
\caption{A generalization of Christofides' algorithm for multi-criteria \mstsp.}
\label{Algo:Christofides}
\end{algorithm}

\begin{theorem}
\label{thm:gastsp}
  For $\gamma \in [\frac 12,1]$, Algorithm~\ref{Algo:Christofides} is a
  randomized $\bigl(\frac{2\gamma^3 + 2 \gamma^2}{3\gamma^2 - 2\gamma +1}
  + \varepsilon\bigr)$-approximation algorithm for multi-criteria \gastsp.
  Its running time is polynomial in the input size and $1/\varepsilon$.
\end{theorem}

\begin{proof}
  The proof consists of two parts. First, we estimate the approximation
  performance, given that all Pareto curves computed are
  $(1+\varepsilon/2)$-approximate Pareto curves. Second, we estimate the success
  probability, \ie, the probability that such a Pareto curve is computed.

  We assume that all Pareto curves that have to be computed during the execution
  of the algorithm were computed successfully, \ie, with an appropriate
  approximation ratio. Let $S'$ be an arbitrary Hamiltonian cycle of $G$.
  Without loss of generality, we assume again that $w_{\min} = 1$. All the
  arguments in the following hold for all $i$.

  There exists a tree $T \in \parmst^\apx$ with
  \[
  w_i(T) \leq \bigl(1+\frac{\varepsilon}2\bigr) \cdot (w_i(S') - 1) .
  \]
  Let $V_\odd$ be the set of vertices of odd degree in $T$, and let $n_\odd$ be
  its cardinality. Note that $n_\odd$ is even, thus perfect matchings exist on
  the complete graph induced by $V_\odd$. Let $n = |V|$ be the number of
  vertices of the whole graph. Let $S_{\odd}$ be the Hamiltonian cycle obtained
  from $S'$ by taking shortcuts. We get two matchings $M_1$ and $M_2$ on
  $V_{\odd}$ from $S_{\odd}$ by putting the edges of $S_{\odd}$ alternately into
  $M_1$ and $M_2$. By $\gamma$-triangle inequality, we have
  \[
      w_i(M_1) + w_i(M_2)
    = w_i(S_{\odd}) \leq w_i(S') - (n-n_\odd) \cdot 2 \cdot (1-\gamma)
  \]
  since every shortcut reduces the weight by at least
  $(1-\gamma) \cdot 2 w_{\min}$. Now,
  $w_i(M_1) \cdot \frac{1-\gamma}{\gamma} \leq w_i(M_2)$ according to
  Lemma~\ref{lem:gastsp}. (Note that we do not make any assumptions whether
  $M_1$ or $M_2$ is the lighter matching.)
  There exists a matching $M \in \parmatch^{\apx}(T)$ with
  \[
  w_i(M) \leq \bigl(1 + \frac{\varepsilon}2 \bigr) \cdot w_i(M_1) .
  \]
  Hence,
  \begin{eqnarray*}
      \frac{w_i(M)}{\gamma} = w_i(M) \cdot \bigl(1+\frac{1-\gamma}{\gamma}\bigr)
  &\leq& \bigl(1+\frac{\varepsilon}2 \bigr) \cdot
         \bigl(w_i(M_1) + w_i(M_2) \bigr) \\
  &\leq& \bigl(1+\frac{\varepsilon}2 \bigr) \cdot \bigl(w_i(S') - 2 \cdot
         (n-n_\odd) \cdot (1-\gamma) \bigr).
  \end{eqnarray*}
  Let $D$ be Eulerian graph obtained by taking the union of the tree $T$ and the
  matching $M$. By the arguments above, we can bound its weight as follows:
  \begin{eqnarray*}
  w_i(D) &  =  & w_i(T) + w_i(M) \\ 
         &\leq & \bigl(1+\frac{\varepsilon}2\bigr)
                 \cdot \bigl(w_i(S') - 1 + \gamma \cdot \bigl(w_i(S') - 2 \cdot
                 (n-n_\odd) \cdot (1-\gamma) \bigr) \bigr) \\
         &  =  & \bigl(1+ \frac{\varepsilon}2\bigr) \cdot \bigl((1+\gamma)
                 \cdot w_i(S') - \bigl(1+2\gamma  \cdot (n-n_\odd)
                 \cdot(1-\gamma) \bigr) \bigr).
  \end{eqnarray*}
  The Eulerian graph $D$ consists of $n-1 + n_\odd/2$ edges, the tour $S$
  constructed from $D$ consists only of $n$ edges. Thus, $n_{\odd}/2 -1$ edges
  are removed and, by $\gamma$-triangle inequality, we have
  \[
    w_i(S) \leq w_i(D) - 2 \cdot (n_\odd/2-1 ) \cdot (1-\gamma)
  = w_i(D) - (1-\gamma) \cdot (n_\odd-2).
  \]
  By combining these inequalities, we obtain
  \begin{eqnarray*}
  w_i(S) &\leq&
  \bigl(1+ \frac{\varepsilon}2\bigr) \cdot \bigl((1+\gamma) \cdot w_i(S') \\
  && \quad - \bigl(1+2 \gamma \cdot (n-n_\odd)\cdot (1-\gamma)
  + (1-\gamma)\cdot (n_\odd-2) \bigr)\bigr) \\
  & = & \bigl(1 +\frac{\varepsilon}2 \bigr) \cdot \bigr(( 1+\gamma)
  \cdot w_i(S') \\
  && \quad - \bigl(\underbrace{1 -2+2\gamma}_{\geq 0} + (1-\gamma) \cdot
  \bigr(\underbrace{2 \gamma}_{\geq 1} \cdot (n-n_\odd) +n_\odd
  \bigr)\bigr)\bigr) \\
  &\leq &\bigl(1 +\frac{\varepsilon}2 \bigr) \cdot \bigl((1+\gamma) \cdot
  w_i(S') -  n \cdot (1-\gamma) \bigr).
  \end{eqnarray*}
  Now we have $w_{\max} \leq \frac{2\gamma^2}{1-\gamma}$ since $w_{\min} = 1$.
  Thus, $w_i(S') \leq \frac{2\gamma^2}{1-\gamma} \cdot n$. We choose
  $\alpha \geq 1$ such that
  $w_i(S') = \frac{2 n \gamma^2}{(1-\gamma) \cdot \alpha}$, which implies
  \begin{eqnarray*}
  \frac{w_i(S)}{w_i(S')} & \leq & \bigl(1+\frac{\varepsilon}2\bigr)
  \cdot\bigl( (1+\gamma) - \frac{n \cdot (1-\gamma)}{w_i(S')} \bigl)
  \;\; \leq \;\;
  \bigl(1+\frac{\varepsilon}2\bigr)
  \cdot \bigl((1+\gamma) -
  \frac{\alpha n\cdot (1-\gamma)}{\frac{2\gamma^2}{1-\gamma} \cdot n}
  \bigr)\\
  &=& \bigl(1+\frac{\varepsilon}2\bigr)
  \cdot\bigl((1+\gamma) - \frac{\alpha (1-\gamma)^2}{2\gamma^2} \bigr) 
  \;\; \leq \;\;
  (1+\gamma) - \frac{\alpha (1-\gamma)^2}{2\gamma^2} + \varepsilon .
  \end{eqnarray*}
  The last inequality holds since
  $(1+\gamma) - \frac{\alpha (1-\gamma)^2}{2\gamma^2} \leq 1+\gamma \leq 2$.

  Since $w_i(S) \leq n \frac{2\gamma^2}{1-\gamma}$, we also have
  $\frac{w_i(S)}{w_i(S')} \leq \alpha$. Thus, the approximation ratio achieved
  is
  \begin{eqnarray*}
  &    & \max_{\alpha\geq 1}\left(\min\left(\alpha, (1+\gamma) -
     \frac{\alpha (1-\gamma)^2}{2\gamma^2} + \varepsilon\right)\right) \\
  &\leq& \frac{2\gamma^3 + 2 \gamma^2}{3\gamma^2 - 2\gamma +1} + \varepsilon
         \cdot \frac{2 \gamma^2}{3 \gamma^2 - 2 \gamma +1}
  \;\; \leq \;\; \frac{2\gamma^3 + 2 \gamma^2}{3\gamma^2 - 2\gamma +1} +
  \varepsilon.
  \end{eqnarray*}
  We obtain the first inequality by observing that
  $(1+\gamma) - \frac{\alpha (1-\gamma)^2}{2\gamma^2} + \varepsilon$ is
  monotonically decreasing in $\alpha$: The maximum of the minimum is therefore
  assumed for $\alpha = (1+\gamma) - \frac{\alpha (1-\gamma)^2}{2\gamma^2} +
  \varepsilon$. The second inequality follows from the fact that
  $\frac{2 \gamma^2}{3 \gamma^2 - 2 \gamma +1} \leq 1$ for
  $\gamma \in [\frac 12, 1]$.

  The analysis so far holds only for $\gamma < 1$ since for $\gamma = 1$,
  division by zero occurs at some points in the analysis. For $\gamma = 1$, we
  obtain a ratio of $2 +\varepsilon =
  \frac{2\gamma^3 + 2 \gamma^2}{3\gamma^2 - 2\gamma +1} + \varepsilon$: We have
  $w(M) \leq \bigl(1+\frac{\varepsilon}2\bigr) \cdot w(S')$ and
  $w(T) \leq \bigl(1+\frac{\varepsilon}2\bigr) \cdot w(S')$, which implies the
  bound.

  What remains to be proved is that the algorithm succeeds with a probability of
  at least $1/2$. First, we observe that if we iterate the randomized
  computation of an approximate Pareto curve, then we do not have to decide
  which set is indeed such an approximate Pareto curve. Instead, we can take the
  union of all solutions computed and remove all dominated solutions of the set
  thus obtained. The only randomization in Algorithm~\ref{Algo:Christofides} is
  in the computation of the approximate Pareto curves for the matching problems.
  The number of curves to be computed is $p$, which is bounded by a polynomial
  of the input size and $1/\varepsilon$. We can achieve a failure probability of
  at most $\frac{1}{2p}$ by performing $\log(2p)$ iterations of the FPRAS for
  the matching problem. The probability that one of the Pareto curve
  computations fails is thus at most $p \cdot \frac{1}{2p} = 1/2$, which
  completes the proof of the theorem.
\end{proof}

We compare the ratios obtained by the two algorithms of this sections and the
cycle cover algorithm of Section~\ref{sec:ccalgo} in
Section~\ref{subsec:comparing}.

%%%%%%%%%%%%%%%%%%%%%%%%%%%%%%%%%%%%
\section{Matchings and Cycle Covers}
\label{sec:cycles}
%%%%%%%%%%%%%%%%%%%%%%%%%%%%%%%%%%%%

A cycle cover of a graph is a spanning subgraph that consists solely of cycles
such that every vertex is part of exactly one cycle. Many approximation
algorithms for the single-criterion TSP are based on cycle covers. These
approximation algorithms usually start by computing an initial cycle cover and
then join the cycles to obtain a Hamiltonian cycle. This technique is called
\emph{subtour patching}~\cite{GilmoreEA:WellSolved:1985}. We show that there
exist FPRASs for multi-criteria cycle cover problems.

\acc, the cycle cover problem in directed graphs, is equivalent to finding
matchings of minimum weight in bipartite graphs (assignment problem). An FPRAS
for the multi-criteria matching problem is also an FPRAS for the multi-criteria
matching problem in bipartite graphs. Hence, multi-criteria \acc\ also admits an
FPRAS.

\begin{theorem}
\label{thm:mcacc}
  There exists an FPRAS for multi-criteria \acc.
\end{theorem}

To show that multi-criteria \scc\ admits an FPRAS, we show that arbitrary graph
factor problems admit FPRASs. Let $G=(V,E)$ be a graph and
$f: V \rightarrow \nat$ be function. A subset $F \subseteq E$ is called an
\bemph{$f$-factor} of $G$ if all vertices $v \in V$ have a degree of exactly
$f(v)$ in the graph $(V,F)$.

Cycle covers of undirected graphs are also known as two-factors since every
vertex is incident to exactly two edges. Thus, they are a special case of graph
factors.

The graph factor problem \bemph{\gfp} is the following minimization problem: An
instance is an undirected graph $G=(V,E)$ with a function
$f: V \rightarrow \nat$ and an edge weight function $w: E \rightarrow \nat$. The
aim is to find an $f$-factor of minimum weight.

To show that multi-criteria \gfp, and thus multi-criteria \scc\ as well, admits
an FPRAS, we exploit Tutte's reduction~\cite{Tutte:FactorTheorem:1954}, which
reduces arbitrary graph factor problems to matchings (matchings are also known
as one-factors since every vertex is incident to exactly one edge of the
matching). We omit a description of the reduction, but refer to Lov{\'a}sz and
Plummer~\cite{LovaszPlummer:Matching:1986} or Tutte~\cite{Tutte:FactorTheorem:1954}
for the details. Overall, we obtain the following result.

\begin{theorem}
  Multi-criteria \gfp\ and multi-criteria \scc\ admit an FPRAS.
\end{theorem}

%%%%%%%%%%%%%%%%%%%%%%%%%%%%%%%%%%%%%%%%%%%%%%
\section{Approximations Based on Cycle Covers}
\label{sec:ccalgo}
%%%%%%%%%%%%%%%%%%%%%%%%%%%%%%%%%%%%%%%%%%%%%%

%%%%%%%%%%%%%%%%%%%%%%%%%%
\subsection{The Algorithm}
\label{subsec:ccgeneral}

The generic outline of a cycle-cover-based algorithm is the following: Start by
computing a cycle cover. Then remove one edge of every cycle. Finally, join the
paths thus obtained to form a Hamiltonian cycle.

Algorithm~\ref{Algo:CCApprox} is our generalization of this algorithm to
multi-criteria TSP. It achieves a constant approximation ratio if the quotient
of the weight of the heaviest edge and the weight of the lightest edge is
bounded.

In this section, we present a general analysis of the approximation ratio of
this algorithm. We will refine the analysis for multi-criteria \gastsp\
(Section~\ref{subsec:ccgastsp}) to get an improved approximation ratio.
Furthermore, we apply the analysis to get approximation results for
multi-criteria \gaatsp\ (Section~\ref{subsec:ccgaatsp}) and \stspot\ and
\atspot\ (Section~\ref{subsec:tspot}). We analyze Algorithm~\ref{Algo:CCApprox}
in terms of the number $\alpha n$ of edges that have to be removed and the
quotient $\beta = w_{\max}/w_{\min}$.

\begin{algorithm}[t]
\begin{algorithmic}[1]
\Input complete graph $G=(V,E)$; $k$ edge weight functions $w_i$
   ($i \in [k]$); $\varepsilon' > 0$
\Output an approximate Pareto curve $\partsp^{\apx}$ to multi-criteria TSP
   (with a probability of at least $1/2$)
\State compute a $(1+\varepsilon')$-approximate Pareto curve $\parcc$ to the
   multi-criteria cycle cover problem on $G$ using the algorithm by
   Papadimitriou and Yannakakis~\cite{PapadimitriouYannakakis:TradeOffs:2000}
\ForAll{cycle covers $C \in \parcc$}
\ForAll{cycles $c$ of $C$}
\State remove one edge of $c$
\EndFor
\State join the paths to form a Hamiltonian cycle $S$
\State add $S$ to $\partsp^{\apx}$
\EndFor
\end{algorithmic}
\caption{An approximation algorithm for multi-criteria TSP based on cycle
   covers.}
\label{Algo:CCApprox}
\end{algorithm}

\begin{lemma}
\label{lem:genericcc}
  Assume that at most $\alpha n$ edges have to be removed from each cycle cover
  and that $\frac{\max_{e \in E} w_i(e)}{\min_{e \in E} w_i(e)} \leq \beta$ for
  all $i \in [k]$.

  Then Algorithm~\ref{Algo:CCApprox} is a randomized
  $\bigl(1 +  \alpha (\beta -1) + \varepsilon\bigr)$ approximation algorithm for
  every $\varepsilon > 0$. Its running-time is polynomial in the input size and
  $1/\varepsilon$.
\end{lemma}

\begin{proof}
  Without loss of generality, let $\min_{e \in E} w_i(e) = 1$ for all
  $i \in [k]$. We run the algorithm with some $\varepsilon'$ that depends on
  $\alpha$, $\beta$, and $\varepsilon$ and that we will specify later on. Let
  $S'$ be an arbitrary Hamiltonian cycle. Then there exists a cycle cover $C$ in
  $\parcc$ with $w(C) \leq (1+\varepsilon') \cdot w(S')$. We obtain a
  Hamiltonian cycle $S$ from $C$ such that
  \[
  w_i(S) \leq w_i(C) +  \alpha n (\beta -1)
  \]
  for all $i \in [k]$. The reason for this is that every edge removed has a
  weight of at least $1$ and every edge added has a weight of at most $\beta$.
  Now we have for all $i\in [k]$
  \begin{eqnarray*}
  \frac{w_i(S)}{w_i(S')} &\leq &(1+\varepsilon') \cdot \frac{w_i(S)}{w_i(C)}
  \;\; \leq \;\; (1+\varepsilon') \cdot
                 \frac{w_i(C) +  \alpha n  (\beta -1)}{w_i(C)} \\
  &\leq & (1+\varepsilon') \cdot \frac{n +  \alpha n  (\beta -1)}{n}
  \;\; = \;\; (1+\varepsilon') \cdot (1 +  \alpha   (\beta -1))  \\
  & \leq & 1 +  \alpha   (\beta -1) + \varepsilon
  \end{eqnarray*}
  for $\varepsilon' \leq \frac{\varepsilon}{1 +  \alpha   (\beta -1)}$, which
  proves the lemma.
\end{proof}

%%%%%%%%%%%%%%%%%%%%%%%%%%%%%%%%%%%%%%%%%%%%%%%%%%%
\subsection{\boldmath Refined Analysis for \gastsp}
\label{subsec:ccgastsp}

From the general analysis (Lemma~\ref{lem:genericcc}), we obtain an
approximation ratio of
$\frac 23 + \frac 23 \cdot  \frac{\gamma^2}{1-\gamma} + \varepsilon$ for
\gastsp. In this section, we present a refined analysis that yields a better
approximation ratio.

\begin{figure}[t]
\centering
\subfigure[Cycle cover, before the patching.]{%
\qquad \includegraphics{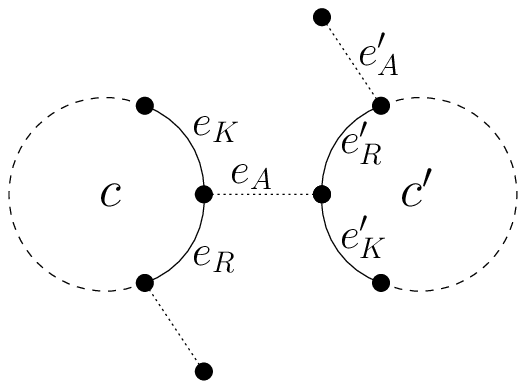} \qquad}
\qquad
\subfigure[Hamiltonian cycle, after the patching.]{%
\qquad \includegraphics{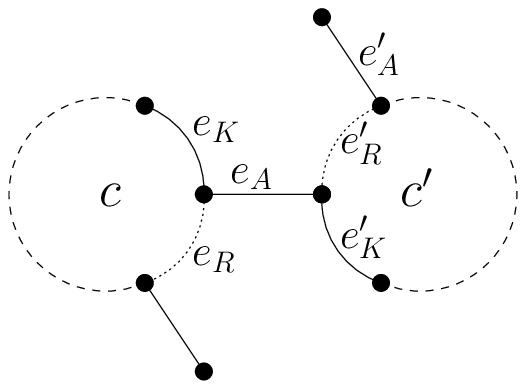} \qquad}
\caption{Two cycles $c$ and $c'$ before and after joining the cycles to a
Hamiltonian cycle. The edges $e_R$, $e_K$, and $e_A$ belong to $c$ while $e_R'$,
$e_K'$, and $e_A'$ belong to $c'$.}
\label{fig:consecutive}
\end{figure}

Consider any cycle $c$ of a cycle cover of $\parcc$. Then there will be an edge
$e_R$ of $c$ that will be removed and an edge $e_A$ adjacent to $e_R$ that will
be added during the joining process. Finally, there exists an edge $e_K$ of $c$
that is adjacent to both $e_R$ and $e_A$ (Figure~\ref{fig:consecutive} shows an
example). Note that while $e_R$ is uniquely determined, once the edges have been
removed and the new edges have been added, the edge $e_A$ is not since there are
two edges that connect $c$ to other cycles of the cycle cover. However, once we
have fixed $e_A$ for one cycle $c$, the corresponding $e_K$ is uniquely
determined, and the $e_A'$ and $e_K'$ of all other cycles $c'$ are also
determined.

By Lemma~\ref{lem:gastsp}, we have
$w_i(e_R) \geq \frac{1-\gamma}{\gamma} \cdot w_i(e_A)$ and
$w_i(e_K) \geq \frac{1-\gamma}{\gamma}  \cdot w_i(e_A)$. All arguments in the
following hold for all weight functions simultaneously. Thus, we restrict
ourselves to considering one fixed weight function $w_i$ for some $i \in [k]$ to
simplify the arguments.

Let $w_R$ be the total weight of edges removed, $w_A$ be the total weight of
edges added, and $w_K$ be the total weight of edges of $C$ and $S$ that are
adjacent to edges added. Then we have $w(C) = w+w_K + w_R$ for some suitably
chosen $w \geq 0$, which is the total weight of all edges not taken into account
so far. Thus,
\[
    \frac{w_i(S)}{w_i(C)}
  = \frac{w + w_K + w_A}{w + w_K + w_R} = 1 + \frac{w_A-w_R}{w + w_K + w_R} = R.
\]
Since $R$ is monotonically decreasing with respect to $w_R$, we obtain
\[
R  \leq 1 + \frac{w_A-\frac{1-\gamma}{\gamma} \cdot w_A}%
                 {w + w_K + \frac{1-\gamma}{\gamma} \cdot w_A}
   =    R'.
\]
Exploiting further that $R'$ is monotonically decreasing in $w_K$, we get
\[
R' \leq 1 + \frac{w_A- \frac{1-\gamma}{\gamma} \cdot w_A}%
                 {w + \frac{1-\gamma}{\gamma} \cdot 2w_A}
   =    1+ \frac{w_A (2\gamma-1)}{\gamma w + (1-\gamma)2 w_A} = R''.
\]
The inequalities $w_A \leq \frac{2 \gamma^2 n}{3 (1-\gamma)}$ and $w \geq n/3$
hold since every cycle has a length of at least three. We exploit the fact that
$R''$ is monotonically increasing with respect to $w_A$ and monotonically
decreasing with respect to $w$:
\[
R'' \leq 1+ \frac{\frac{2 \gamma^2 n}{3 (1-\gamma)} \cdot
         (2\gamma-1)}{\frac{\gamma n}3+ (1-\gamma) \cdot 2
         \cdot \frac{2 \gamma^2 n}{3 (1-\gamma)}}
    =    1+ \frac{\frac{2 \gamma^2}{1-\gamma} \cdot (2\gamma-1)}%
                 {\gamma+ 4 \gamma^2}
    =    \frac{1+\gamma}{1+3 \gamma - 4 \gamma^2}.
\]

We run the algorithm with some $\varepsilon' > 0$ that depends on $\gamma$. We
will specify $\varepsilon'$ in a moment. Let $S'$ be an arbitrary Hamiltonian
cycle and $C \in \parcc$ be a cycle cover with
$w_i(C) \leq (1+\varepsilon') \cdot w(S')$ for all $i \in [k]$. Let $S$ be the
Hamiltonian cycle obtained from $C$. Then
\[
w_i(S) \leq \bigl(1+\varepsilon') \cdot
            \frac{1+\gamma}{1+3 \gamma - 4 \gamma^2} \cdot w_i(S') .
\]
For a given $\varepsilon > 0$, we choose $\varepsilon'$ such that
$\varepsilon' \cdot \frac{1+\gamma}{1+3 \gamma - 4 \gamma^2} \leq \varepsilon$.
The set $\partsp^{\apx}$ is a
$\bigl(\frac{1+\gamma}{1+3 \gamma - 4 \gamma^2} + \varepsilon\bigr)$-approximate
Pareto curve with a probability of at least $1/2$, which implies the following theorem.

\begin{theorem}
\label{thm:gastspcc}
  For $\gamma \in [\frac 12, 1)$, Algorithm~\ref{Algo:CCApprox} is a randomized
  $\bigl(\frac{1+\gamma}{1+3 \gamma - 4 \gamma^2} +
  \varepsilon\bigr)$-approximation algorithm for all $\varepsilon > 0$. Its
  running-time is polynomial in the input size and $1/\varepsilon$.
\end{theorem}

In Section~\ref{subsec:comparing}, we compare the approximation ratios of the
cycle cover algorithm for \gastsp\ to the tree doubling and Christofides'
algorithm.

%%%%%%%%%%%%%%%%%%%%%%%%%%%%%%%%%%%%%%%%%%%%%%%%%%%%%%%%%%%%
\subsection{\boldmath The Cycle Cover Algorithm for \gaatsp}
\label{subsec:ccgaatsp}

For multi-criteria \gaatsp, our algorithm yields a constant factor approximation
if $\gamma < \frac{1}{\sqrt 3}$ since $w_{\max}/w_{\min}$ is bounded from above
by $\frac{2\gamma^3}{1-3\gamma^2}$ for such $\gamma$. For larger values of
$\gamma$, this ratio can be unbounded.

\begin{lemma}[Chandran and Ram~\cite{ChandranRam:Parameterized:2007}]
\label{lem:charam}
  Let $\gamma \in [1/2,1)$. Let $G=(V,E)$ be a directed complete 
  graph, and let $w: E \rightarrow \nat$ be an edge weight function
  satisfying $\gamma$-triangle inequality.
  Let $w_{\min} = \min_{e \in E} w(e)$ and $w_{\max} = \max_{e \in E} w(e)$.

  If $\gamma < 1/\sqrt 3$, then $\frac{w_{\max}}{w_{\min}} \leq
  \frac{2\gamma^3}{1 - 3 \gamma^2}$. If $\gamma \geq 1/\sqrt 3$, then
  $\frac{w_{\max}}{w_{\min}}$ can be unbounded.
\end{lemma}

By combining Lemma~\ref{lem:genericcc} and Lemma~\ref{lem:charam}, we obtain the
following result.

\begin{theorem}
\label{thm:gaatspcc}
  For $\gamma < 1/\sqrt 3$, Algorithm~\ref{Algo:CCApprox} is a randomized
  $\bigl(\frac 12  +  \frac{\gamma^3}{1 - 3 \gamma^2}  +
  \varepsilon\bigr)$-approximation algorithm for \gaatsp. Its running-time is
  polynomial in the input size and $1/\varepsilon$.
\end{theorem}

\begin{figure}[t]
\centering
\includegraphics{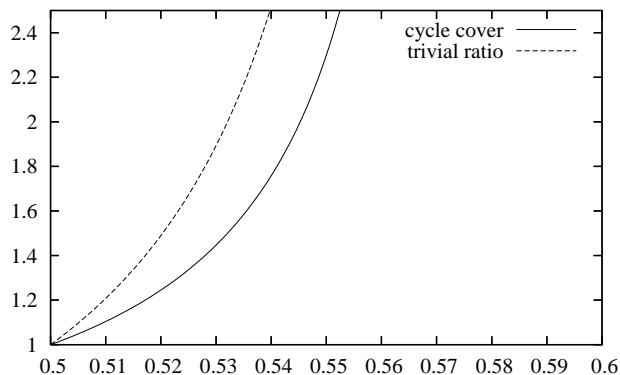}
\caption{The approximation ratio of Algorithm~\ref{Algo:CCApprox} achieved for
\gaatsp\ subject to $\gamma$ compared to the trivial approximation ratio of
$w_{\max} / w_{\min}$.}
\label{fig:atsp}
\end{figure}

Figure~\ref{fig:atsp} shows the approximation ratio achieved for multi-criteria
\gaatsp\ subject to $\gamma$ and compared to the trivial ratio of $w_{\max}/w_{\max}$.

We leave as an open problem to generalize the analysis to larger values of
$\gamma$. However, it seems to be hard to find a constant factor approximation
for $\gamma = 1$, \ie, for multi-criteria \matsp, since this would immediately
yield a constant factor approximation for single-criterion \matsp.

%%%%%%%%%%%%%%%%%%%%%%%%%%%%%%%%%%%%%%%%%
\subsection{TSP with Weights One and Two}
\label{subsec:tspot}

Now we analyze the cycle cover algorithm for multi-criteria TSP with weights one
and two. For both \stspot\ and \atspot, we have $\beta = 2$, \ie,
$w_{\max}/w_{\min} = 2$. Furthermore, for \stspot, we have $\alpha \leq 1/3$,
while we only have $\alpha \leq 1/2$ in case of \atspot. The approximation ratio
follows by exploiting  Lemma~\ref{lem:genericcc}.

Note that the edge weights and thus the objective functions are polynomially
bounded for \stspot\ and \atspot. Thus, we can compute a Pareto curve of cycle
covers instead of only a $(1+\varepsilon)$-approximate Pareto curve. This
implies that we do not have an additional $\varepsilon$ in the approximation
ratios in the following theorems.

\begin{theorem}
\label{thm:stspot}
  Algorithm~\ref{Algo:CCApprox} is a randomized $4/3$-approximation algorithm
  for multi-crite\-ria \stspot. Its running-time is polynomial.
\end{theorem}

\begin{theorem}
\label{thm:atspot}
  Algorithm~\ref{Algo:CCApprox} is a randomized $3/2$-approximation algorithm
  for multi-crite\-ria \atspot. Its running-time is polynomial.
\end{theorem}

%%%%%%%%%%%%%%%%%%%%%%%%%%%%
\section{Concluding Remarks}
%%%%%%%%%%%%%%%%%%%%%%%%%%%%

%%%%%%%%%%%%%%%%%%%%%%%%%%%%%%%%%%%%%%%%%%%%%%%
\subsection{Comparing the Approximation Ratios}
\label{subsec:comparing}

Let us compare the approximation ratios for \gastsp\ achieved by the tree
doubling algorithm (Algorithm~\ref{Algo:Tree}), Christofides' algorithm
(Algorithm~\ref{Algo:Christofides}), and the cycle cover algorithm
(Algorithm~\ref{Algo:CCApprox}).

\begin{figure}[t]
\centering
\includegraphics{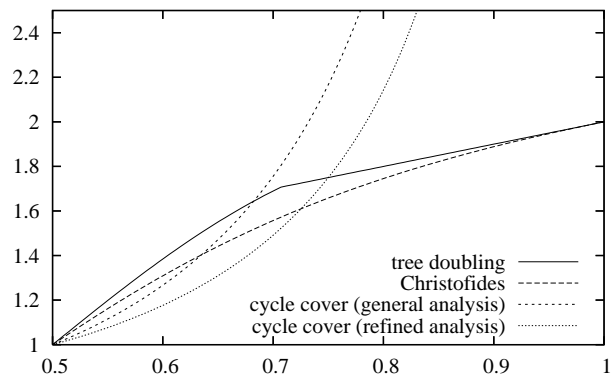}
\caption{Approximation ratios subject to $\gamma$ achieved by the tree doubling
algorithm (Algorithm~\ref{Algo:Tree}), Christofides' algorithm
(Algorithm~\ref{Algo:Christofides}), and the cycle cover algorithm
(Algorithm~\ref{Algo:CCApprox}, Section~\ref{sec:ccalgo}), for which both the
ratio obtained from the general analysis (Section~\ref{subsec:ccgeneral}) and
from the refined analysis (Section~\ref{subsec:ccgastsp}) are shown.}
\label{fig:algos}
\end{figure}

\begin{figure}[t]
\centering
\includegraphics{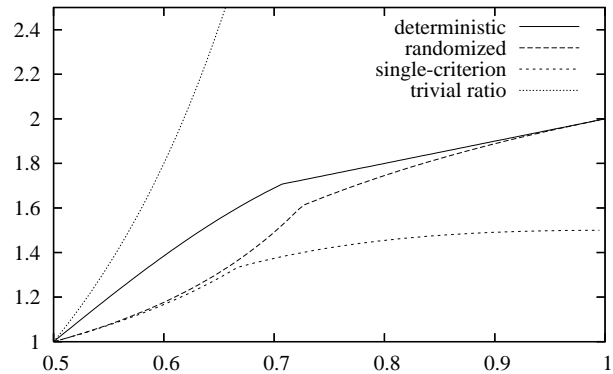}
\caption{Approximation ratios subject to $\gamma$. The deterministic ratio is
achieved by the tree doubling algorithm. Combining Christofides' and the cycle
cover algorithm yields the randomized ratio. For comparison, the current ratio
for single-criterion \gastsp\ and the trivial ratio $w_{\max}/w_{\min}$ are also
shown.}
\label{fig:best}
\end{figure}

Figure~\ref{fig:algos} shows the approximation ratios achieved by these
algorithms subject to~$\gamma$. Figure~\ref{fig:best} shows the approximation
ratios achieved deterministically (by the tree doubling algorithm) and
randomized (by a combination of Christofides' and the cycle cover algorithm).
The ratios are compared to the trivial ratio of $w_{\max}/w_{\min}$ and to the
currently best known approximation ratio for single-criterion \gastsp. Note that
in particular for small values of $\gamma$, our algorithms for multi-criteria
\gastsp\ come close to achieving the ratio of the best algorithms for
single-criterion \gastsp.

%%%%%%%%%%%%%%%%%%%%%%%%%%
\subsection{Open Problems}
\label{subsec:conclusions}

Our approximation algorithm for multi-criteria \gaatsp\ works only for
$\gamma < 1/\sqrt{3}$. Thus, we are interested in finding constant factor
approximation algorithms also for $\gamma \geq 1/\sqrt{3}$, which exist for all
$\gamma < 1$ for single-criterion
\gaatsp~\cite{BlaeserEA:ATSP:2006,ChandranRam:Parameterized:2007}

The cycle-cover-based algorithm for Max-TSP, where Hamiltonian cycles of maximum
weight are sought, does not seem to perform well for multi-criteria Max-TSP. The
reason for this is that the approximation algorithms for Max-TSP that base on
cycle covers usually contain a statement like ``remove the lightest edge of
every cycle''. While this works for single-criterion TSP, the term ``lightest
edge'' is not well-defined for multi-criteria traveling salesman problems. We
are particularly curious about the approximability of multi-criteria Max-TSP.

%%%%%%%%%%%%%%%%%%%%%%%%%%
\section*{Acknowledgments}
%%%%%%%%%%%%%%%%%%%%%%%%%%

We thank Jan Arpe for valuable comments.

%%%%%%%%%%%%%%%%%%%%%%%%%%%%%%%%%%%%%%%%%%%%%%%%%%%%%%%%%%%%%%%%%%%

\end{document}